\documentclass[a4paper, 10 pt, conference]{hsmr} 
\usepackage[utf8]{inputenc}
\IEEEoverridecommandlockouts                       
\usepackage{mathtools}
\usepackage{geometry}
\usepackage[labelsep=space]{caption}
\usepackage{newtxmath,newtxtext}
\usepackage{authblk}
\usepackage{pgfplots}
\usepackage{graphicx}
\usepackage[colorlinks,urlcolor=blue]{hyperref} 
\usepackage{newtxtext,newtxmath}

\pgfplotsset{compat=newest}

\graphicspath{{./images/}}
\DeclareGraphicsExtensions{.pdf,.png,.jpg}

\makeatletter
\let\NAT@parse\undefined
\makeatother
\usepackage[numbers,compress]{natbib}
\bibpunct{[}{]}{,}{n}{}{;}% to get correct punctuation for bibliography
\definecolor{orange}{rgb}{1,0.2,0}

\usepackage{subfig}

\title{
\LARGE \bf
Pose Estimation for Intra-cardiac Echocardiography Catheter via AI-Based Anatomical Understanding
}

\author{\large Jaeyoung Huh$^{}$}
\author{\large Ankur Kapoor$^{}$} 
\author{\large Young-Ho Kim$^{}$}

\affil{\normalsize\textit{$^{}$Digital Technology \& Innovation, Siemens Healthineers, Princeton, NJ, USA,} \\ 
\small\textit{(jaeyoung.huh, ankur.kapoor, young-ho.kim)@siemens-healthineers.com} \vspace{-10pt}}

\begin{document}

\maketitle
\thispagestyle{empty}
\pagestyle{empty}

\section*{INTRODUCTION}

Intra-cardiac Echocardiography (ICE) is a widely used real-time, high-resolution imaging modality that provides critical visualization of cardiac structures from within the heart. It plays a vital role in both Electrophysiology (EP) procedures and Structural Heart Disease (SHD) interventions, allowing clinicians to perform complex cardiac procedures with improved precision and safety.

In EP procedures, accurate catheter localization is crucial. CARTO (Biosense Webster Inc., USA) integrates Electro-Magnetic (EM)-based tracking with anatomical mapping to enable precise ICE catheter navigation. In contrast, SHD interventions typically lack position-tracking systems, requiring operators to manually adjust ICE imaging to explore key anatomical structures such as the left atrial appendage (LAA), pulmonary veins (PV), and atrial septum. This process relies heavily on operator experience and may require frequent adjustments.

EM-based tracking relies on static anatomical maps and is susceptible to magnetic interference, causing position drift or inaccuracies. In ICE-only procedures, the absence of tracking demands extensive manual adjustments, increasing procedural complexity. These limitations highlight the need for an alternative approach that directly leverages ICE images rather than relying solely on external tracking systems.

Recent advancements in AI-driven ICE catheter navigation have introduced autonomous view recovery and AI-based assistance to reduce operator workload and improve procedural efficiency. Automated view recovery systems allow clinicians to bookmark critical imaging views and return to them at the push of a button, facilitating efficient and repeatable ICE imaging during interventions\,\cite{kim2022automated}. Additionally, an AI-driven view guidance system has been proposed to assist users in navigating ICE imaging without requiring extensive expertise\,\cite{huh2025viewguidance}. These AI-based systems aim to enhance ICE manipulation efficiency, particularly for less experienced operators.

We propose an anatomy-aware pose estimation system that determines ICE catheter position and orientation using only ICE images, eliminating the need for external sensors. This enables relative pose estimation, expressing catheter position in relation to key anatomical landmarks rather than a pre-defined map.

This method enhances ICE navigation intuition, making it easier to reach target anatomical structures. It is particularly beneficial for position-tracking-free procedures and can complement existing mapping systems like CARTO, providing real-time anatomical understanding for improved localization.

\section*{MATERIALS AND METHODS}
\vspace{-5pt}

\begin{figure}[t]
	\centering 
	\includegraphics[width= 0.45\textwidth]{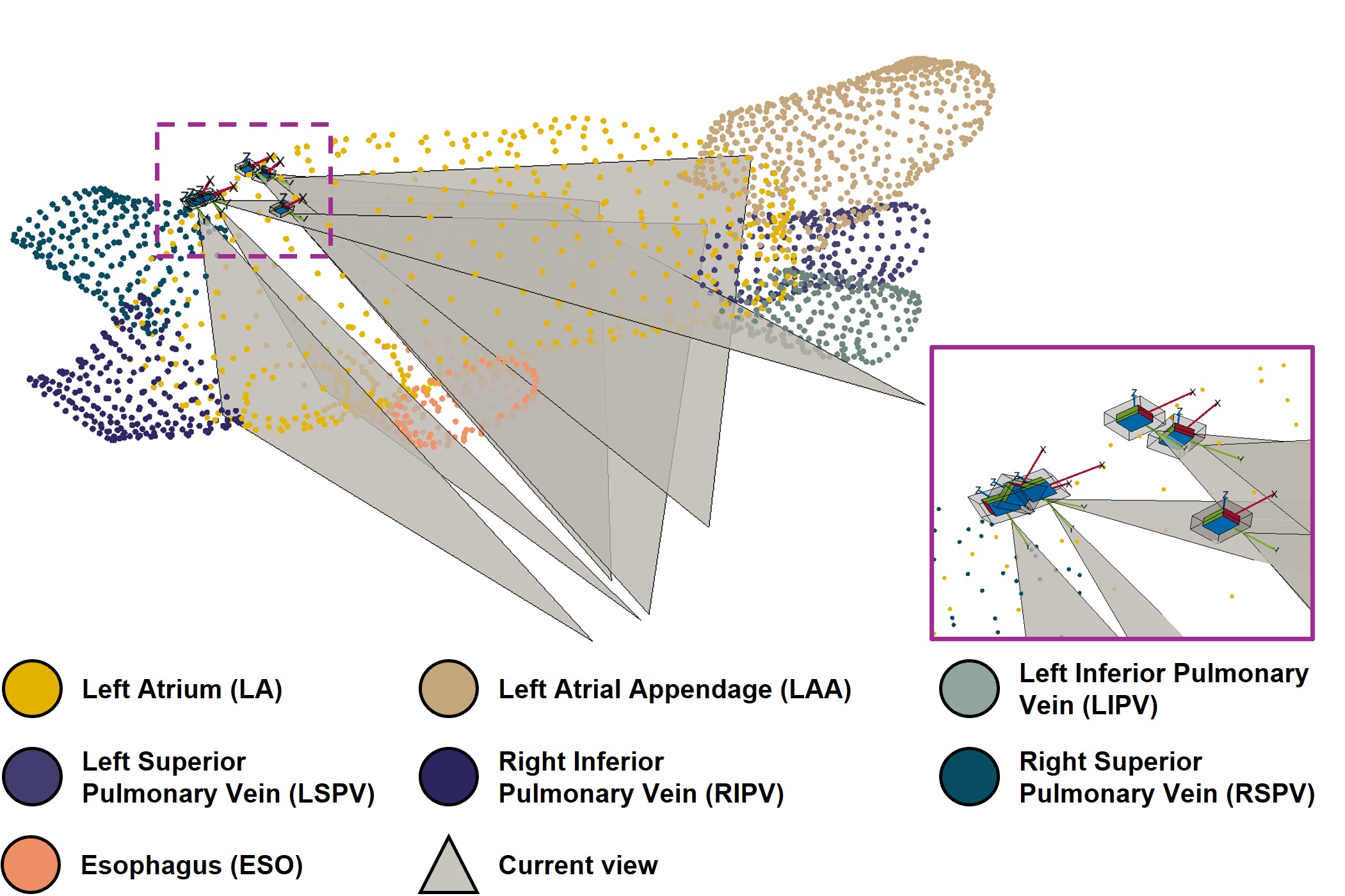}
	\caption{Each colored dot represents the left atrium and pulmonary vein structure of a single subject. The gray fan indicates the current ICE image view, serving as the input to our model. The red, blue, and green arrows within the purple box illustrate the predicted orientation orientation and position of the imaging plane, which is expressed in anatomy-relative coordinates.}
    \label{fig:dataset}
    \vspace{-10pt}
\end{figure}

To estimate the ICE catheter's pose purely from ICE images, we trained a deep learning model that learns spatial relationships between ICE catheter pose and their corresponding anatomical view images. 

\vspace{-5pt}
\subsection{Dataset}
\vspace{-5pt}
We utilized a dataset collected from a well-established clinical environment and processed using the CARTO mapping system\,\cite{jnj2023carto}. The dataset includes ICE images paired with their corresponding position and orientation information, as well as cardiac mesh data. A total of 851 subjects were included, with the dataset split into 793 for training, 25 for validation, and 33 for testing. Since each dataset was collected independently, they each had their own world coordinate system. To ensure consistency across samples, all position and orientation data were normalized relative to the center of each left atrium (LA) mesh.
The transformation from the world coordinate-based transducer state ($S_{w}^{tr}$) to the center of the LA mesh-based state ($S_{m}^{tr}$) was achieved using a transformation matrix ($T_{w}^{m}$). This normalization process can be expressed as follows:
\begin{eqnarray}
    S_{m}^{tr} = (T_{w}^{m})^{-1} * S_{w}^{tr}.
\end{eqnarray}
The dataset includes 18,305 training samples, with 813 for validation and 823 for testing. Position and orientation data were normalized relative to the left atrium center for consistency.

\vspace{-5pt}
\subsection{Training details}
\begin{figure}[t!]
	\centering 
	\includegraphics[width= 0.45\textwidth]{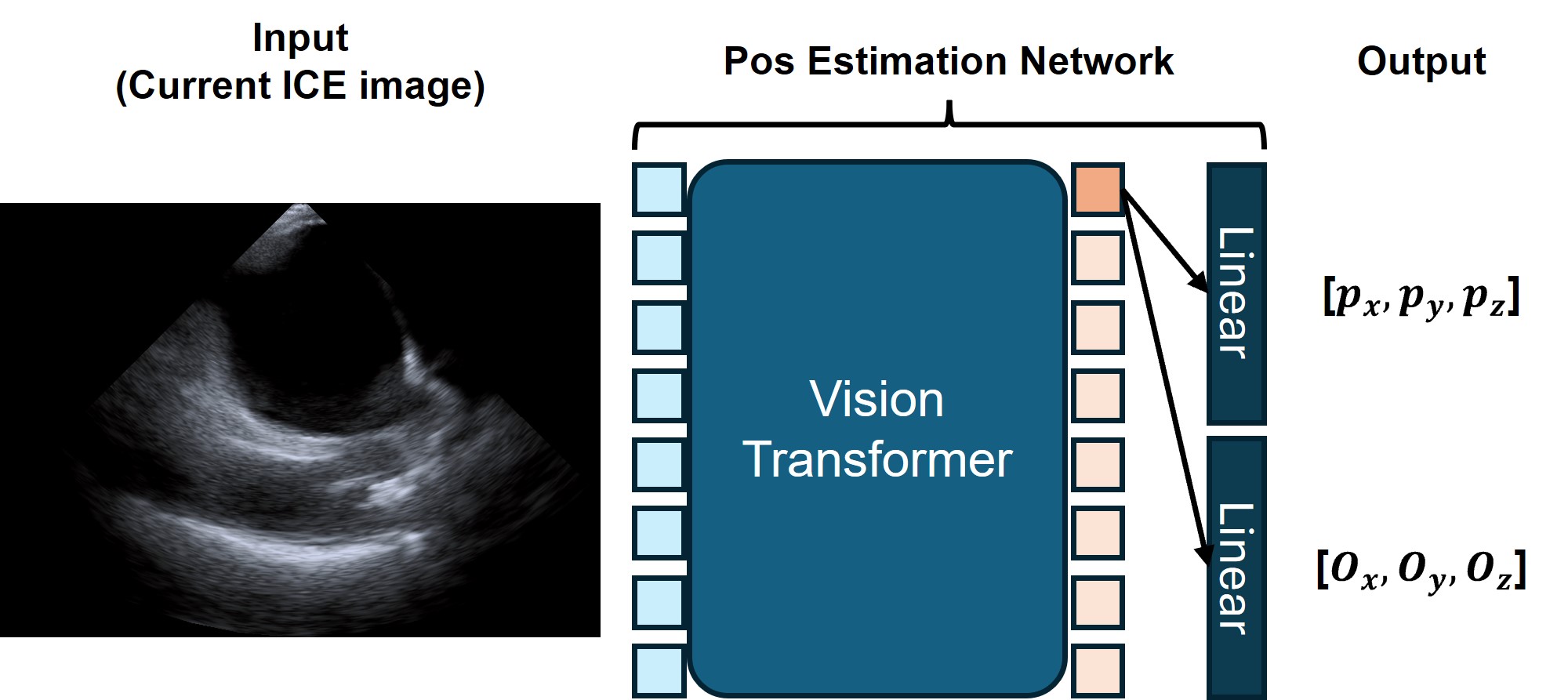}
	\caption{Network architecture: The input ICE image is patchified and fed into the ViT. The \textit{[CLS]} token from the output is then passed through separate linear layers to estimate position and orientation independently.}  
    \label{fig:network}
    \vspace{-10pt}
\end{figure}

\vspace{-5pt}
We trained a Vision Transformer (ViT)-based network $\mathcal{N}$ to capture global spatial relationships within ICE images\,\cite{dosovitskiy2020image}. Each ICE image is divided in $16 \times 16$ patches, embedded into 768 dimensional vectors with positional encoding. A class token (\textit{[CLS]}) is appended, and the sequence is processed through the transformer network. The \textit{[CLS]} token output is passed through two separate linear layers to predict position $\hat{P}$ and orientation $\hat{O}$ as $\hat{P}, \hat{O} = \mathcal{N}(I)$. The overview network is illustrated in Figure\,\ref{fig:network}. The model was trained using Mean Squared Error (MSE) loss, with a total loss function: 
\begin{eqnarray}
    l_{total}   = l_{mse}(\hat{P}, P) + \lambda * l_{mse}(\hat{O}, O), 
\end{eqnarray} 
where $\lambda = 2$ balances position and orientation errors. 

The entire network was trained for 140 epochs with a batch size of 16. Training was implemented in PyTorch and conducted on an NVIDIA A100 GPU.

% These patches are flattened and linearly projected into embedding vectors, each of size 768. Positional encoding is then added to the embedding vectors to incorporate spatial information, and a class token (\textit{[CLS]}) is appended to represent the overall image. The resulting input sequence is defined as $F=[CLS, e_{1}, e_{2},...,e_{N}]$, where $e_{1}$ represents the embedding vector of the first patch, and $N$ denotes the total number of patches. This input sequence is fed into the transformer network. Two separate linear layers are added to estimate position and orientation information independently. The \textit{[CLS]} token from the transformer's output is passed through each linear layer, yielding predictions for position and orientation. The overview architecture is illustrated in Figure\,\ref{fig:network}. The entire process is summarized as follows:
% \begin{eqnarray}
%     \hat{P}, \hat{O} = N(I),\\
%     l_{total}   = l_{mse}(\hat{P}, P) + \lambda * l_{mse}(\hat{O}, O),
% \end{eqnarray}
% where the outputs of the two linear layers are position (P) and orientation (O), while the predicted values are denoted as $\hat{P}, \hat{O}$, respectively. The $N$ is the proposed network. To train the model, We used mean-squared error (MSE) loss function, represented as $l_{mse}$. The total loss is a weighted combination of the position and orientation losses, where $\lambda$ is the weighting parameter. In this experiment, $\lambda=2$ was used. The entire training process was implemented in PyTorch and conducted on a single NVIDIA A100 GPU.

% \subsection{Dataset}
% To achieve the goal of image-based pose estimation, we utilized a dataset collected from a well-established clinical environment and processed using the CARTO mapping system (Biosense Webster Inc., USA) \cite{jnj2023carto}. The dataset includes ICE images paired with their position and orientation information, as well as corresponding cardiac mesh data. A total of 851 subjects were included, with the dataset split into 793 for training, 25 for validation, and 33 for testing. Since each dataset was collected independently, they each have their own world coordinate system. To ensure consistency across datasets, we normalized all position and orientation data relative to the center of each left atrium (LA) mesh. The transformation from the world coordinate-based transducer state ($S_{w}^{tr}$) to the center of the LA mesh-based state ($S_{m}^{tr}$) was achieved using a transformation matrix ($T_{w}^{m}$). This normalization process can be expressed as follows:
% \begin{eqnarray}
%     S_{m}^{tr} = (T_{w}^{m})^{-1} * S_{w}^{tr}.
% \end{eqnarray}
% Each subject contributed multiple images visualizing various cardiac structures, resulting in a total of 18305 paired datasets comprising images along with their corresponding position and orientation information for the training process. The validation and test processes utilized 813 and 823 paired datasets, respectively.

\begin{figure}
	\centering 
	\includegraphics[width= 0.45\textwidth]{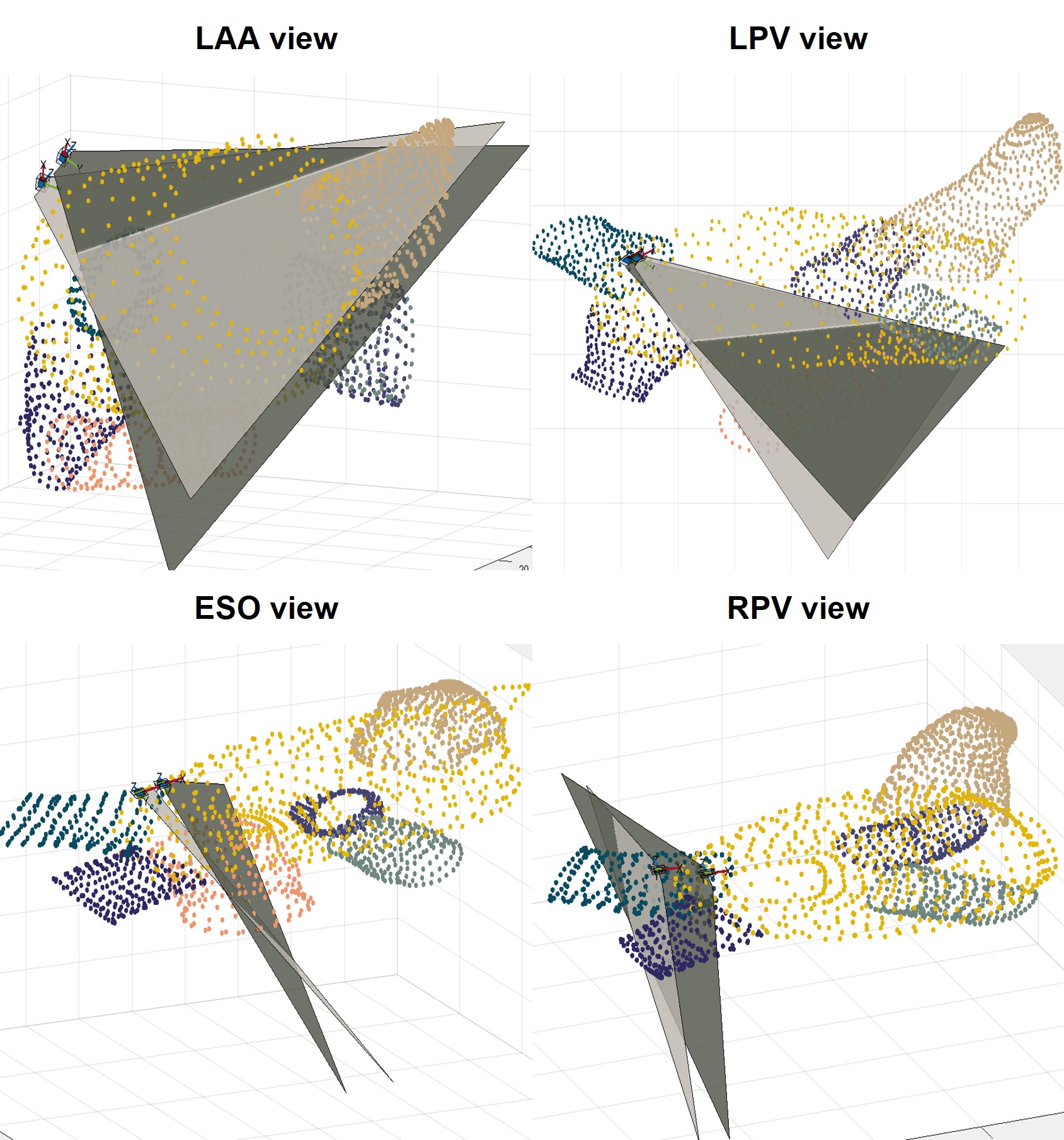}
	\caption{Representative results of the proposed method: The top-left case visualizes the LAA view, the top-right case visualizes the LPV view, the bottom-left case visualizes the ESO view, and the bottom-right case visualizes the RPV view.}
    \label{fig:result}
    \vspace{-10pt}
\end{figure}

\vspace{-5pt}
\section*{RESULTS}
\vspace{-5pt}
We validated our method through qualitative and quantitative evaluations using the paired dataset.

\noindent{\bf Qualitative Assessment:} Figure \ref{fig:result} illustrates the alignment between the predicted imaging fan (dark green) and the target fan (gray) within a 3D cardiac mesh. The model accurately visualizes key structures, such as the left atrial appendage (LAA) and right pulmonary vein (RPV), with minimal positional and orientational errors.

\noindent{\bf Quantitative Evaluation:} Table \ref{table:error} summarizes the errors, with an average positional error of 9.48 mm and orientation errors of (16.13, 8.98, 10.47) degrees across the x-, y-, and z-axes. These results confirm the high accuracy of our method in estimating catheter position and orientation.

% For the qualitative assessment, we examined whether the predicted imaging fan was well-aligned with the target fan and its corresponding cardiac mesh. We visualized both the target and predicted fans along with the mesh in a 3D space, with representative results shown in Figure \ref{fig:result}. In these visualizations, the gray fan represents the original imaging plane, while the dark green fan denotes the predicted one. In the top-left of Figure \ref{fig:result}, the original fan captures the left atrial appendage (LAA) structure, and the predicted fan accurately visualizes the same structure with minimal positional and orientational error. Similarly, in the bottom-right of Figure \ref{fig:result}, the original and predicted fans effectively visualize the right pulmonary vein (RPV) structure, demonstrating the reliability of our method.

% For quantitative evaluation, we computed the positional and orientational errors between the predicted and target values. The mean errors across all test samples are summarized in Table \ref{table:error}. The average positional error is within 9.48 mm, while the orientation errors for the x-, y-, and z-axes are (16.13, 8.98, 10.47) degrees, respectively. These results demonstrate the model's high accuracy in estimating the original position and orientation.

\begin{table}[h!]
    \centering
    \caption{Prediction error}
	\resizebox{0.45\textwidth}{!}{
		\begin{tabular}{|c|c|c|}
			\hline 
              & Position & Orientation \\ \hline
		Mean error   & 9.48 [mm] & (16.13, 8.98, 10.47) [degree] \\
            Std      & 5.96 [mm] & (42.31, 9.47, 14.81) [degree] \\ \hline
		\end{tabular}
	}
	\label{table:error}
\end{table}

\vspace{-5pt}
\section*{DISCUSSION}
\vspace{-5pt}
We presented a ViT-based method for ICE catheter pose estimation using only ICE images. Both qualitative and quantitative evaluations confirm its accuracy in predicting position and orientation.
Our system can function independently or alongside position sensors, enhancing accuracy, safety, and procedural efficiency in ICE-based interventions.

\vspace{-5pt}
\section*{DISCLAIMER}
\vspace{-5pt}
{
The concepts and information presented in this paper are based on research results that are not commercially available. Future availability cannot be guaranteed.
}

%\nocite{*}
{
\bibliography{references_hamlyn}
}

\end{document}